\documentclass[11pt,a4paper]{article}
\pdfoutput=1

\usepackage[left=0.8in,right=0.8in,top=1.0in,bottom=1.2in]{geometry}

\usepackage{mathrsfs,graphicx,rotating,amsmath,amsfonts,mathtools,booktabs,amssymb,wasysym,caption}
\usepackage{cite}
\usepackage{hyperref}
\usepackage{authblk}
\usepackage{slashed}
\usepackage[table,xcdraw,dvipsnames]{xcolor}
\usepackage{graphicx}
\usepackage{bbold}
\usepackage[utf8x]{inputenc}
\usepackage[english]{babel}
\usepackage{multirow,multicol}
\usepackage{epstopdf}
\usepackage{bbm}
\usepackage{changepage}
\usepackage{appendix}
\usepackage{libertine}
\usepackage{braket}
\usepackage{wasysym}
\usepackage{empheq}
\usepackage{cancel}
\usepackage{enumitem}
\usepackage{mathrsfs}
\usepackage{lmodern}
\usepackage{tabularx}
\usepackage{multicol}
\usepackage{multirow}
\usepackage{color}
\usepackage{mathtools}
\usepackage{verbatim}
\usepackage{arydshln}
\usepackage{url}
\usepackage[normalem]{ulem}
\usepackage{youngtab}

\newcommand{\gsim}{\lower.7ex\hbox{$\;\stackrel{\textstyle>}{\sim}\;$}}
\newcommand{\lsim}{\lower.7ex\hbox{$\;\stackrel{\textstyle<}{\sim}\;$}}

\title{\textbf{Phenomenology of Minimal Flavour Deconstruction at the lowest new scale}}
 
\author{Riccardo Barbieri\thanks{riccardo.barbieri@sns.it}}

\affil{Scuola Normale Superiore, Piazza dei Cavalieri 7, 56126 Pisa, Italy}

\begin{document} 
\maketitle

\abstract{ 
I describe in detail the phenomenology of a d=4 flavour-non-universal gauge theory, based on Ref. \cite{Barbieri:2023qpf}  , where global accidental symmetries  control  the pattern of the Yukawa couplings of the charged fermions and, at the same time, the deviations from the Standard Model of  the relevant flavour observables. Special attention is payed to the phenomenology at the lowest new scale, the mass $m_{Z_{23}}$ of a neutral vector boson, in the MultiTeV.
}

\tableofcontents 

\section{Introduction}
\label{Intro}

As one of the most successful theories of a quadrant of nature ever formulated, the Standard Model (SM) of Particle Physics leaves open, not surprisingly,  a few fundamental questions, both of observational and of structural origin. 
Among the second ones, a striking fact is that  the non vanishing SM particle masses, particularly the fermion masses as well as their intergenerational mixing angles, are all taken from experiments without any understanding whatsoever of their values or even of their patterns. As such, the {\it flavour puzzle} is a strong motivation for the existence of New Physics (NP) beyond the SM. The question is at which scale, $\Lambda_{NP}$, since this is crucial to properly orient its search, if it exists at all, or perhaps to tie it to the {\it hierarchy problem} as close as possible to the ElectroWeak (EW) scale.

As well known, a blind implementation of a SM Effective Field Theory analysis confines the scale of new physics associated with flavour and CP violation to PeVs or more\cite{Bona:2024bue}. At the same time, from the observed masses and mixings, the quark Yukawa couplings, $Y_f = U_L^{f+}Y_f^{diag} U_R^f, f=u,d$,      
exhibit an approximate $U(2)_q\times U(2)_u\times U(2)_d$ symmetry\cite{Barbieri:2011ci}, provided $[U_L^{u,d}]_{i\neq j}\lesssim [V_{CKM}]_{i\neq j}$ and $[U_R^{u,d}]_{i\neq j}\lesssim [U_L^{u,d}]_{i\neq j}$. This raises  an interesting question: could $U(2)_q\times U(2)_u\times U(2)_d$ emerge in a suitable extension of the SM as an accidental symmetry accounting for the pattern of the Yukawa couplings and, at the same time, allowing for the scale of new physics  at much lower values than PeVs, closer to the EW scale?

A tentative positive answer to this question has been put forward in several recent papers, all based on flavour-non-universal gauge extensions, dubbed {\it flavour deconstruction}, of the SM\cite{Davighi:2022bqf,Davighi:2023iks,FernandezNavarro:2023rhv,Davighi:2023evx,Covone:2024elw}. Here I analyse in detail the phenomenology of a model, contained in Ref.  \cite{Barbieri:2023qpf} , based on the gauge 
 group
\begin{equation}
G = SU(3)\times SU(2)\times U(1)_Y^{[3]} \times U(1)_{B-L}^{[12]} \times U(1)_{T_{3R}}^{[2]} \times U(1)_{T_{3R}}^{[1]},
\label{eq:G}
\end{equation}
where $SU(3)$ and $SU(2)$ act universally on the three fermion families, as in the SM, whereas the  $U(1)$ groups 
act non-universally only on one or two families, as indicated by the corresponding superscripts. Given the explicit particle content of the model, recalled in the next Section, the symmetry breaking cascade of the $U(1)$ factors down to the SM $U(1)_Y$ gives indeed rise to an approximate, global accidental symmetry, extended to leptons, 
\begin{equation}
U(2)^5 \equiv U(2)_q\times U(2)_u\times U(2)_d\times U(2)_l \times U(2)_e
\label{U5}
\end{equation}
 and, at a higher scale, to a subgroup of it, 
 \begin{equation}
 U(1)^3\equiv U(1)_{u_1}\times U(1)_{d_1}\times U(1)_{e_1},
 \label{U3}
 \end{equation}
 which accounts for the smallness of the first generation masses of the charged fermions.

\section{The model defined}
\label{Mod}

For ease of the reader I recall the particle content of the model~\cite{Barbieri:2023qpf}. Table \ref{tab:SSB} lists the scalar fields, responsible for the breaking of the $U(1)$ factors of the gauge group in two steps, by $\langle\sigma\rangle >> \langle\phi, \chi\rangle $, 
as well as for EW symmetry breaking by the two doublets $H_{u,d}$, distinguished by a softly broken $Z_2$ symmetry which makes them couple to the up-type quarks/neutrinos  and to the down-type quarks/charged leptons respectively.
As to the fermions,  other than  the usual three sixteen-plets of chiral fermions, whose quantum numbers are  self-explanatory, they include  three generations of Vector-Like (VL) $SU(2)$-singlets 
shown in Table~\ref{Table:VLFA}.  

\begin{figure}[t]
\centering
\includegraphics[clip,width=.95\textwidth]{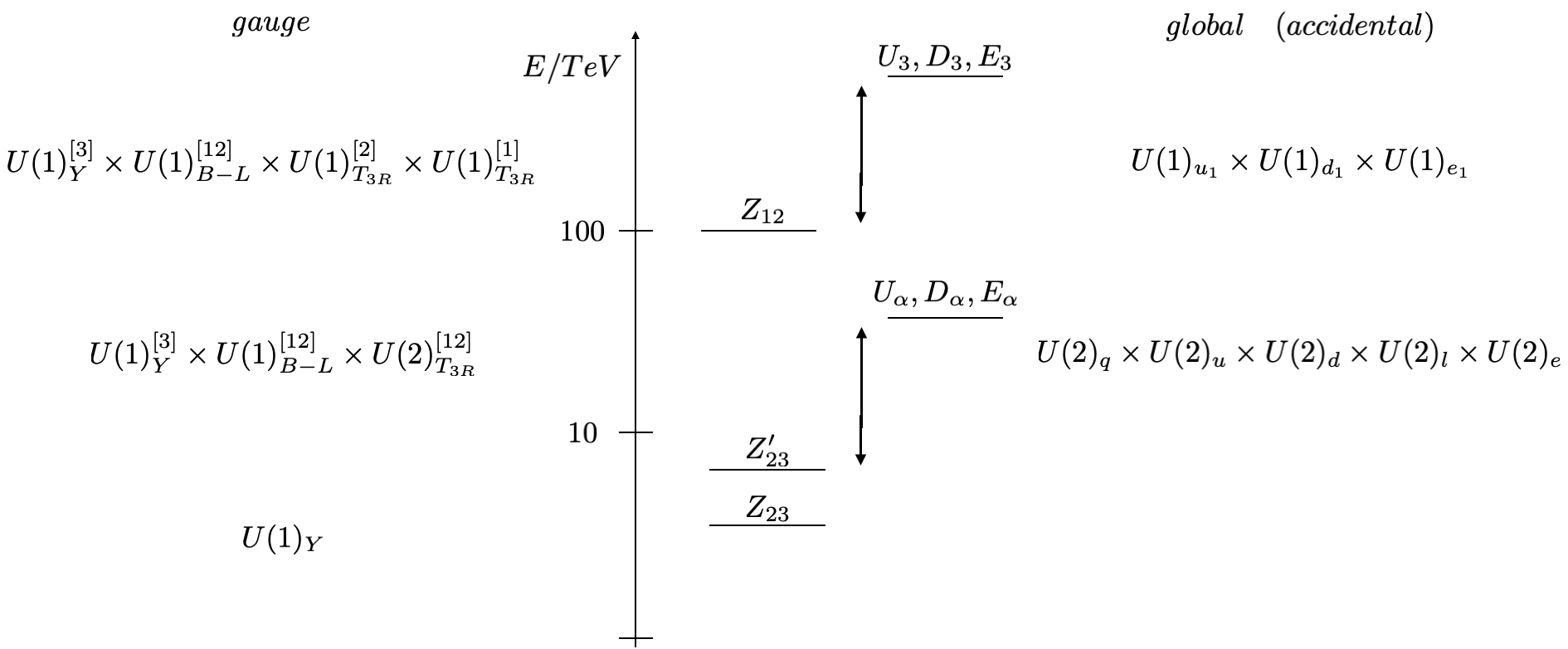}
\caption{\small  Overall representation of the model. On the left and on the right are the chains of gauge ($SU(3)\times SU(2)$ left understood) and global symmetries (Universal $U(1)_B\times U(1)_L$ left understood) respectively. In the centre are the masses of the new particles: neutral vectors, $Z_{23}, Z_{23}', Z_{12}$, and VL $SU(2)$-singlet fermions, $U_i, D_i, E_i, i=1,2,3$.  The vertical lines with two arrows denote the separation of scales which controls the breaking of the global symmetries.}
\label{fig:Overall}
\end{figure}

The overall picture of the model is represented in Fig.~\ref{fig:Overall}, with an indication of the relevant scales inferred from the preliminary analysis of Ref.~\cite{Barbieri:2023qpf}. In the following I pay special attention to the phenomenology associated with the lowest new scale occurring in the model, the mass  $m_{Z_{23}}$ of the lightest neutral vector boson.

\begin{table}[t]
 $$\begin{array}{c||c|c|c|c|c}
 {\rm Field} &  U(1)_Y^{[3]} & U(1)_{B-L}^{[12]} & U(1)_{T_{3R}}^{[2]}& U(1)_{T_{3R}}^{[1]}& SU(3)\times SU(2)\\ \hline
 H_{u,d} & -1/2& 0 & 0 & 0 &(\bold{1},\bold{2}) \\ \hline \hline
 \chi^q   &-1/6& 1/3 & 0 &0 & (\bold{1},\bold{1}) \\ \hline 
 \chi^l & 1/2& -1 & 0 &0 & (\bold{1},\bold{1}) \\ \hline 
 \phi & 1/2& 0 & -1/2& 0&(\bold{1},\bold{1}) \\ \hline \hline
\sigma &0 & 0 & 1/2 & -1/2  & (\bold{1},\bold{1}) \\ \hline 
\end{array}$$
\caption{\small  Scalar fields responsible for the symmetry-breaking pattern of the gauge group.}
\label{tab:SSB}
\end{table}

 \begin{table}[t]
 $$\begin{array}{cc|c|c|c|c|c}
& & U(1)_Y^{[3]} & U(1)_{B-L}^{[12]} & U(1)_{T_{3R}}^{[2]}& U(1)_{T_{3R}}^{[1]}& SU(3)\times SU(2)\\ \hline
\multirow{3}{*}{  $\begin{array}{c} {\rm light\  VL} \\  (\alpha=1,2) \end{array}$   }
& U_\alpha & 1/2& 1/3 & 0 &0 &(\bold{3},\bold{1}) \\ \cline{2-7}
& D_\alpha & -1/2& 1/3 & 0 & 0&(\bold{3},\bold{1}) \\ \cline{2-7}
& E_\alpha & -1/2& -1 & 0 & 0&(\bold{1},\bold{1}) \\ 
 \hline\hline
\multirow{3}{*}{ heavy\ VL   }  
&U_3 & 0& 1/3 & 1/2 &0 &(\bold{3},\bold{1}) \\ \cline{2-7}
&D_3 & 0& 1/3 & -1/2 &0 &(\bold{3},\bold{1}) \\ \cline{2-7}
&E_3 & 0& -1 & -1/2 & 0&(\bold{1},\bold{1}) \\ \hline
\end{array}$$
\caption{\small  Vector-like fermions }
\label{Table:VLFA}
\end{table}

The full set of Yukawa-like couplings and fermion mass terms is determined by the transformation properties of these fields. For example in the up-quark sector $(i=1,2, \alpha = 1,2)$
\begin{eqnarray}
\mathcal{L}^u_Y &=&  (y^u_3\, \bar{q}_{3} u_{3} H_u +  y^u_{i\alpha}\,  \bar{q}_{i} U_{\alpha}  H_u 
+  y^{\chi_u}_{\alpha}\,   \bar{U}_{\alpha} u_{3}  \chi^q + y^{\phi_u}_{\alpha 2}\,  \bar{U}_{\alpha} u_{2} \phi  
+ y^{\phi_u}_{\alpha 3}\,    \bar{U}_{R\alpha} U_{L3}  \phi \nonumber \\
&&  +\,  \hat{y}^{\phi_u}_{\alpha 3}\,    \bar{U}_{L\alpha} U_{R3}  \phi  + y^{\sigma_u}_1\,  \bar{U}_{3} u_{1} \sigma+ {\rm h.c.} ) + M_{U_3}\, \bar{U}_{3}U_3 + M_{U_\alpha} \,   \bar{U}_{\alpha}{U}_{\alpha}
\label{LYA}
\end{eqnarray}
where, unless specified, the chirality component is left understood since non ambiguous $(q \equiv q_L, u \equiv u_R)$.
In the down-quark  sector everything is similar, but for the replacement  $H_u \to H_d$, as in the charged-lepton sector where also $\chi^q \to \chi^l$.
As indicated in Fig.~\ref{fig:Overall}, apart from exact universal $U(1)_B\times U(1)_L$, Eq.~(\ref{LYA})  makes clear that in the limit $M_{U_\alpha}, M_{U_3}\rightarrow \infty $ the model exhibits an accidental global $U(2)^5$, Eq.~(\ref{U5}), reduced to $U(1)^3$, Eq.~(\ref{U3}), in the $M_{U_3}\rightarrow \infty $ limit. For finite $M_{U_\alpha}, M_{U_3}$, the breaking of these symmetries is controlled by three parameters, 
\begin{equation}
\epsilon_\phi =  \frac{ \langle  \phi  \rangle}{M_\alpha},\quad  \epsilon_\chi =  \frac{ \langle  \chi  \rangle}{M_\alpha}, \quad \epsilon_\sigma =  \frac{ \langle  \sigma  \rangle}{M_3},
\end{equation}
represented in Fig.~\ref{fig:Overall} by the vertical lines with two arrows.

By integrating out the heavy VL fermions one gets   the true Yukawa couplings of the massless fermions. In the up-type case one obtains
 \begin{equation}
 Y_u  \approx  \begin{pmatrix}
  y^u_{1\alpha} \hat{y}^{\phi_u}_{\alpha 3}  y_1^{\sigma_u} \epsilon_\sigma \epsilon_{\phi }  &    -y^u_{1\alpha} y^{\phi_u}_{\alpha 2} \epsilon_\phi  & -y^{u}_{12}  y^{\chi_u}_{2} \epsilon_\chi   \\
   y^u_{2\alpha}  \hat{y}^{\phi_u}_{\alpha 3}  y_1^{\sigma_u} \epsilon_\sigma \epsilon_{\phi } & -y^u_{2\alpha} y^{\phi_u}_{\alpha 2} \epsilon_\phi  &   -y^{u}_{22}  y^{\chi_u}_{2}  \epsilon_\chi  \\
 \approx 0 &  \approx 0 & y^u_3
 \end{pmatrix}
   \label{eq:YuA}
 \end{equation}
 and similarly for $Y_{d,e}$. For $v_u/v_d \approx 10$ and $\epsilon_\phi \approx \epsilon_\chi \approx \epsilon_\sigma \approx 0.05\div 0.1$ the charged fermion masses and quark mixings are described by all the Yukawa couplings $y$'s in Eq.~(\ref{LYA}) in the $0.1\div 1$ range. In particular the matrix elements of $U_L^{u,d.e}$ have similar size to the matrix elements of 
$V_{CKM}$ (with $U_L^u U_L^{d+} = V_{CKM}$) and $[U_R^{u,d,e}]_{i\neq j} << [U_L^{u,d,e}]_{i\neq j} $.
 Needless to say one is far from having reduced the number of free parameters in the SM - the ultimate goal to solve the flavour puzzle - but the pattern of the charged fermion masses clearly emerges  as a consequence of the approximate accidental global symmetries.

\section{Vector mass eigenstates}
\label{Vme}

The kinetic term of any of the matter fields in Tables \ref{tab:SSB} and  \ref{Table:VLFA} incorporates the covariant derivative 
\begin{equation}
D_\mu \equiv \partial_\mu -i (g_3 Y^{[3]}A_{3\mu} + g_B\frac{(B-L)^{[12]}}{2}A_{B\mu} + g_2 T_{3R}^{[2]}A_{2\mu}
+  g_1 T_{3R}^{[1]}A_{1\mu})
\label{Dmu}
\end{equation}
which defines, with canonically normalised gauge boson fields, all the gauge coupling constants.   The flavour-universal $SU(3)\times SU(2)$ contributions are left understood.

Before ElectroWeak symmetry breaking, in the basis
\begin{equation}
\mathcal{V}_\mu =
 \begin{pmatrix}
  A_{3\mu}  \\
A_{B\mu} \\
A_{T\mu} \\
C_{T\mu}
 \end{pmatrix}, \quad A_{T\mu} \equiv \frac{g_2 A_{1\mu} +g_1 A_{2\mu}}{\sqrt{g_1^2 + g_2^2}} ,
  \quad C_{T\mu} \equiv   \frac{g_1 A_{1\mu}  -g_2 A_{2\mu}}{\sqrt{g_1^2 + g_2^2}} ,
 \end{equation}
the vector mass matrix is
\begin{equation}
M^2 =
 \begin{pmatrix}
 g_3^2 (A +B) & -g_3g_B A & -g_3g_T B & g_3\hat{g}_2 B \\
-g_3g_B A  & g_B^2 A & 0 & 0 \\
-g_3g_T B & 0 & g_T^2 B & -g_T \hat{g}_2  B\\
g_3\hat{g}_2 B & 0 & -g_T \hat{g}_2  & \hat{g}_2^2B + (g_1^2+g_2^2) C
 \end{pmatrix},
 \end{equation}
 where
 \begin{equation}
g_T =\frac{g_1g_2}{\sqrt{g_1^2 + g_2^2}}, \quad \hat{g}_2 = \frac{g_2^2}{\sqrt{g_1^2 + g_2^2}}
 \end{equation}
 and
  \begin{equation}
A =\frac{1}{2}(|\langle\chi^l\rangle |^2 + \frac{1}{9}|\langle\chi^q\rangle |^2 ), \quad B =\frac{1}{2}|\langle\phi\rangle|^2 , \quad C =\frac{1}{2}|\langle\sigma\rangle|^2.
 \end{equation}

For $\langle\sigma\rangle >> \langle\phi, \chi\rangle $, up to negligible corrections of order $B/C= (\langle\phi\rangle / \langle\sigma\rangle)^2$, $C_{T\mu} $ is already a mass eigenstate, which couples to matter fields via the covariant derivative term
\begin{equation}
D_\mu(C) = 
 -i  g_T (T_{3R}^{[2]}- T_{3R}^{[1]}) C_{T\mu}.
\label{DmuC}
\end{equation}
The three other mass eigenstates 
 have simple explicit expressions for $g_3 >> g_B, g_T$.  They are
 \begin{equation}
\mathcal{V}_\mu^{ph} =
 \begin{pmatrix}
  B_{\mu}  \\
Z_{23\mu} \\
Z_{23\mu}' \\
 \end{pmatrix} = R
  \begin{pmatrix}
  A_{3\mu}  \\
A_{B\mu} \\
A_{T\mu} 
 \end{pmatrix},
 \label{Aphys}
 \end{equation}
where
\begin{equation}
R=
 \begin{pmatrix}
\epsilon_B \epsilon_T (1+b)/r & \epsilon_T/r & \epsilon_B/r \\
(\epsilon_B^2-b\epsilon_T^2)/r & \epsilon_B/r & -\epsilon_T/r \\
1 &- \epsilon_B & - b \epsilon_T
 \end{pmatrix},
 \quad \epsilon_{B,T} =\frac{g_{B,T}}{g_3}\frac{1}{1+b},
 \quad r =\sqrt{ \epsilon_B^2+ \epsilon_T^2},
 \quad b = \frac{B}{A},
 \label{R}
 \end{equation}
and they have respective masses
\begin{equation}
m_B^2 =0,  \quad m_{Z_{23}}^2 = (g_B^2+ g_T^2) A\frac{b}{1+b},  \quad m_{Z_{23}'}^2 = g_3^2 A (1+b).
\end{equation}

By expressing the covariant derivative, Eq.~(\ref{Dmu}), in terms of the physical vectors, one finds how they interact with the matter fields. For the $B_\mu$-dependent term, without any approximation, it is
\begin{equation}
D_\mu (B) = -i \frac{g_Bg_T}{\sqrt{g_B^2+g_T^2}} B_\mu Y, \quad Y= Y^{[3]}+ \frac{(B-L)^{[12]}}{2} + T_{3R}^{[12]},
\end{equation}
as expected, so that $g' =\frac{g_Bg_T}{\sqrt{g_B^2+g_T^2}}$ is the SM hypercharge. Altogether, introducing the angle $\alpha$ and the proper combination $a_{23}$,
\begin{equation}
\frac{g_B}{g_T} = \tan\alpha \equiv t_\alpha,
\quad a_{23} = \frac{1-b t_\alpha^2}{ t_\alpha (1+b)},
\end{equation}
the relevant covariant derivative terms become
\begin{equation}
D_\mu (B, Z_{23},  Z_{23}') = -i [g'  B_\mu Y + g'  Z_{23 \mu}(a_{23}Y^{[3]} + \frac{1}{t_\alpha}\frac{(B-L)^{[12]}}{2}- t_\alpha  T_{3R}^{[12]}) + g_3 Z_{23 \mu}' Y^{[3]}].
\label{Dmu-phys}
\end{equation}
Note for later use that, in terms of $b, t_\alpha$, it is 
\begin{equation}
\frac{g_3^2}{g'^2} \frac{m_{Z_{23}}^2}{m_{Z_{23}'}^2}= \frac{b(1+t_\alpha^2)^2}{t_\alpha^2 (1+b)^2}\equiv a_3
\label{a3}
\end{equation}

\section{$Z- Z_{23}, Z_{23}'$ mixing}
\label{ZZmix}

ElectroWeak symmetry breaking gives rise to mixing between the $Z$-boson and the heavier $Z_{23}, Z_{23}'$, which corrects in a relevant way both the mass of the $Z$ and its couplings  to the standard fermions.

The relevant Lagrangian term, for $H_u$ or  for $H_d$, is
\begin{equation}
\mathcal{L}_H^{gauge}=|(\partial_\mu -ig \vec{T}\cdot\vec{W}_\mu -ig_3 Y^{[3]} A_{3\mu})H|^2
\end{equation}
which, upon use of Eq.s~(\ref{Aphys},\ref{R}) for $A_{3\mu}$ and recalling that $Y^{[3]}_H=-1/2$, leads to the mass term for the vector bosons
\begin{equation}
\mathcal{L}_m=\frac{1}{2}g^2v^2 W_\mu^+W_\mu^- + \frac{1}{4}v^2 (-\sqrt{g^2+g'^2} Z_\mu + g' a_{23} Z_{23 \mu} +g_3 Z_{23 \mu}' )^2,
\quad v^2 = v_u^2 + v_d^2.
\end{equation}
To a sufficient level of approximation the $Z- Z_{23}, Z_{23}'$ mixings appearing in this equation, supplemented with the mass terms for the $Z_{23}, Z_{23}'$ bosons discussed in the previous Section, are eliminated by the replacements
\begin{equation}
Z_{23 \mu} \rightarrow Z_{23 \mu} + s_W a_{23} \frac{m_Z^2}{m_{Z_{23}}^2}Z_\mu , \quad
Z_{23 \mu}' \rightarrow Z_{23 \mu}' + s_Wg_3 \frac{m_Z^2}{m_{Z_{23}'}^2}Z_\mu, \quad
s_W = \frac{g'}{\sqrt{g^2 + g'^2}},
\label{shifts}
\end{equation}
which correct both the mass and the couplings of the $Z$-boson. 
The corrected square mass is
\begin{equation}
m_Z^2 = \frac{1}{2}(g^2+ g'^2) v^2 [1-s_W^2(a_{23}^2 \frac{m_Z^2}{m_{Z_{23}}^2} + \frac{g_3^2}{g'^2}\frac{m_Z^2}{m_{Z_{23}'}^2})]
\end{equation}
or, upon use of Eq.~(\ref{a3}), 
\begin{equation}
m_Z^2 = \frac{1}{2}(g^2+ g'^2) v^2 [1-s_W^2\frac{m_Z^2}{m_{Z_{23}}^2}(a_{23}^2  + a_3)]
\label{deltamZ}
\end{equation}
Similarly from Eq.s~(\ref{Dmu-phys},\ref{a3},\ref{shifts}) one obtains a correction to the couplings of the $Z$ 
\begin{equation}
\delta D_\mu (Z) =  -i Z_\mu g's_W\frac{m_Z^2}{m_{Z_{23}}^2}[(a_{23}^2 +a_3)  Y^{[3]}+a_{23}(\frac{1}{t_\alpha}\frac{(B-L)^{[12]}}{2}- t_\alpha  T_{3R}^{[12]})]]
\label{Dmu-Z}
\end{equation}

\section{Tree level phenomenology at the lowest new scale $m_{Z_{23}}$}

In the context of the model, the general question about the scale of new physics referred to in the Introduction becomes which is the lower bound allowed by current experiments on the mass of the lightest new particle, $m_{Z_{23}}$. As already indicated in Ref.~\cite{Barbieri:2023qpf}  and in previous work in similar contexts,  the relevant data include flavour observables, ElectroWeak Precision Tests and high $p_T$ events at LHC. At least at tree level Sect.s \ref{Vme}  and \ref{ZZmix} show that all these data depend on three parameters, $m_{Z_{23}}, t_\alpha, b$, and, in the case of flavour, as well on the element $[U_L^f]_{3i}$ of the unitary matrices that diagonalises $Y_f$ on the left side. In the following, rather than an overall fit, I discuss individually a few examples, since this is more relevant to understand their significance and to follow the evolution of the data.

\subsection{Corrections to the $Z$ mass}

\begin{figure}[t]
\centering
\includegraphics[clip,width=.95\textwidth]{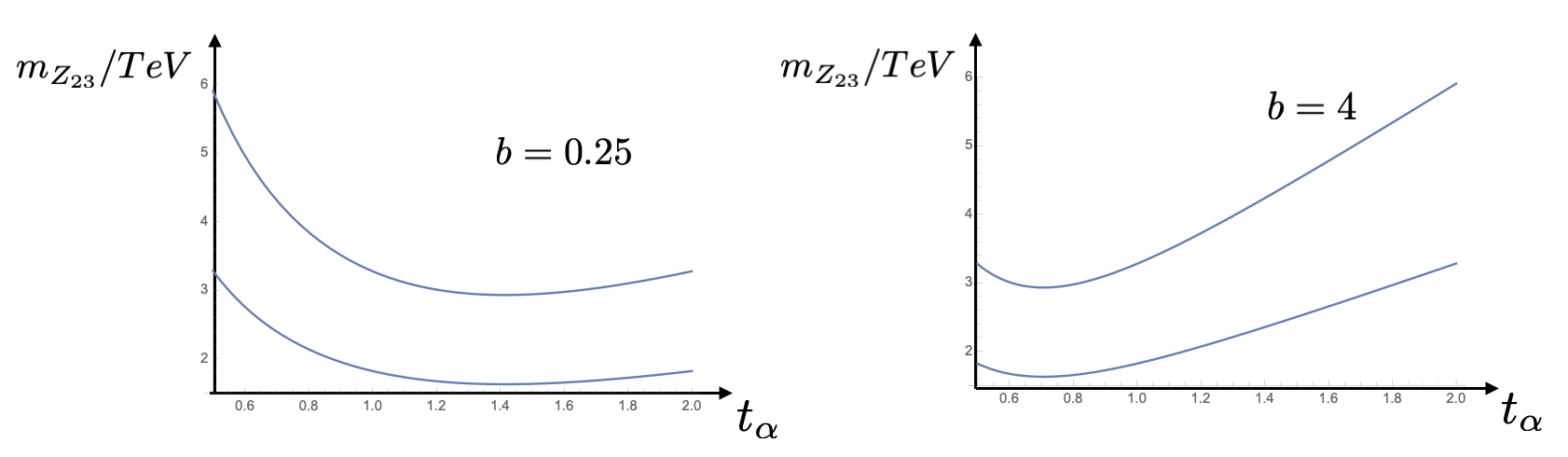}
\caption{\small  Range of values for $m_{Z_{23}}$ that would reproduce the deviation of  $m_Z(SM)$ from $m_Z(exp)$ in 
the $1 \sigma$ interval of Eq.~(\ref{delta-rho}). On the left $b= 0.25$, on the right $b= 4$. }
\label{fig:deltam_Z}
\end{figure}

Taking Eq.~(\ref{deltamZ})   as the single (main) correction to the EW precision observables, 
\begin{equation}
\frac{\delta m_Z^2}{m_Z^2} \equiv s_W^2\frac{m_Z^2}{m_{Z_{23}}^2}(a_{23}^2  + a_3)
\label{a23a3}
\end{equation}
is given by
\begin{equation}
\frac{\delta m_Z^2}{m_Z^2} = (\frac{m_Z(SM)}{m_Z(exp)})^2 -1=(38\pm 20)\cdot 10^{-5}
\label{delta-rho}
\end{equation}
where
$m_Z(exp)= 91.1875\pm 0.0021$  GeV is the experimental value of the $Z$-mass and $m_Z(SM)= 91.2047\pm 0.0088$ GeV~\cite{deBlas:2021wap} is the result of the overall EW precision fit without the inclusion of $m_Z(exp)$\footnote{
The same result is obtained by including in the full  EW precision fit the parameter $\epsilon_1 $\cite{Altarelli:1991fk}, to be identified with $\frac{\delta m_Z^2}{m_Z^2}$, or a  parameter $\rho_0=1+ \epsilon_1$ as in Ref.~\cite{ParticleDataGroup:2022pth}. The recent result by CDF \cite{CDF:2022hxs} on $m_W$ is not included in the fit in view of the significant tension with other measurements of the same quantity.
}.
From Eq.~(\ref{a23a3}) the range of values for $m_{Z_{23}}$ that would reproduce the $1 \sigma$ interval in Eq.~(\ref{delta-rho}) is represented in Fig.~\ref{fig:deltam_Z}
for $t_\alpha = 0.5\div 2$ - so that $g_{B.T} <1$ and $g_3 >> g_{B.T}$ makes sense - and $b=0.25\div 4$, i.e. $\langle\chi\rangle / \langle\phi\rangle =0.5\div 2$.

\subsection{$pp\rightarrow  l^+l^-$ for $l=e,\mu$ }

In the model under consideration the Drell-Yan process $pp\rightarrow  l^+l^-$ for $l=e,\mu$ is predominantly due to $u\bar{u}, d\bar{d}\rightarrow  l^+l^-$ mediated by the $Z_{23}$ exchange through its couplings to any light fermion $f$, neglecting small off diagonal terms (see below)
\begin{equation}
\mathcal{L}_{Z_{23}}=g' Z_{23 \mu} (g_L^f\bar{f}_L\gamma^\mu f_L + g_R^f\bar{f}_R\gamma^\mu f_R).
\label{gVA}
\end{equation}
 In the narrow width approximation all the model dependence is included in the coefficients~\cite{Carena:2004xs,Accomando:2010fz}
 \begin{equation}
c_u = g'^2(g^{u 2}_L + g^{u 2}_R) Br_{Z_{23}}(l^+l^-), \quad c_d = g'^2(g^{d 2}_L + g^{d 2}_R) Br_{Z_{23}}(l^+l^-),
\end{equation}
which are related to the cross section $\sigma_{l^+l^-} \equiv \sigma (pp\rightarrow Z_{23} \rightarrow l^+l^-)$ via
 \begin{equation}
c_u w_u + c_d w_d = \frac{6}{\pi} \sigma_{l^+l^-}
\end{equation}
and $w_{u,d}$ are the $u \bar{u}, d \bar{d}$ parton luminosities.

\begin{table}[t]
 $$\begin{array}{c||c|c|c}
 m_{Z_{23}}/TeV &  \sigma_{l^+l^-}^{lim}/ fb & w_u/fb &w_d/fb  \\ \hline
 5  &  0.017 & 16  &4.1 \\ \hline 
 5.5 &  0.017 &  6.0 & 1.9\\ \hline 
 6 & 0.017  & 2.4 & 0.9
\end{array}$$
\caption{\small  Inputs used to get the bounds on $m_{Z_{23}}$ in Fig. \ref{fig:c_ud}
}
\label{Tab:data}
\end{table}

\begin{figure}[t]
\centering
\includegraphics[clip,width=.55\textwidth]{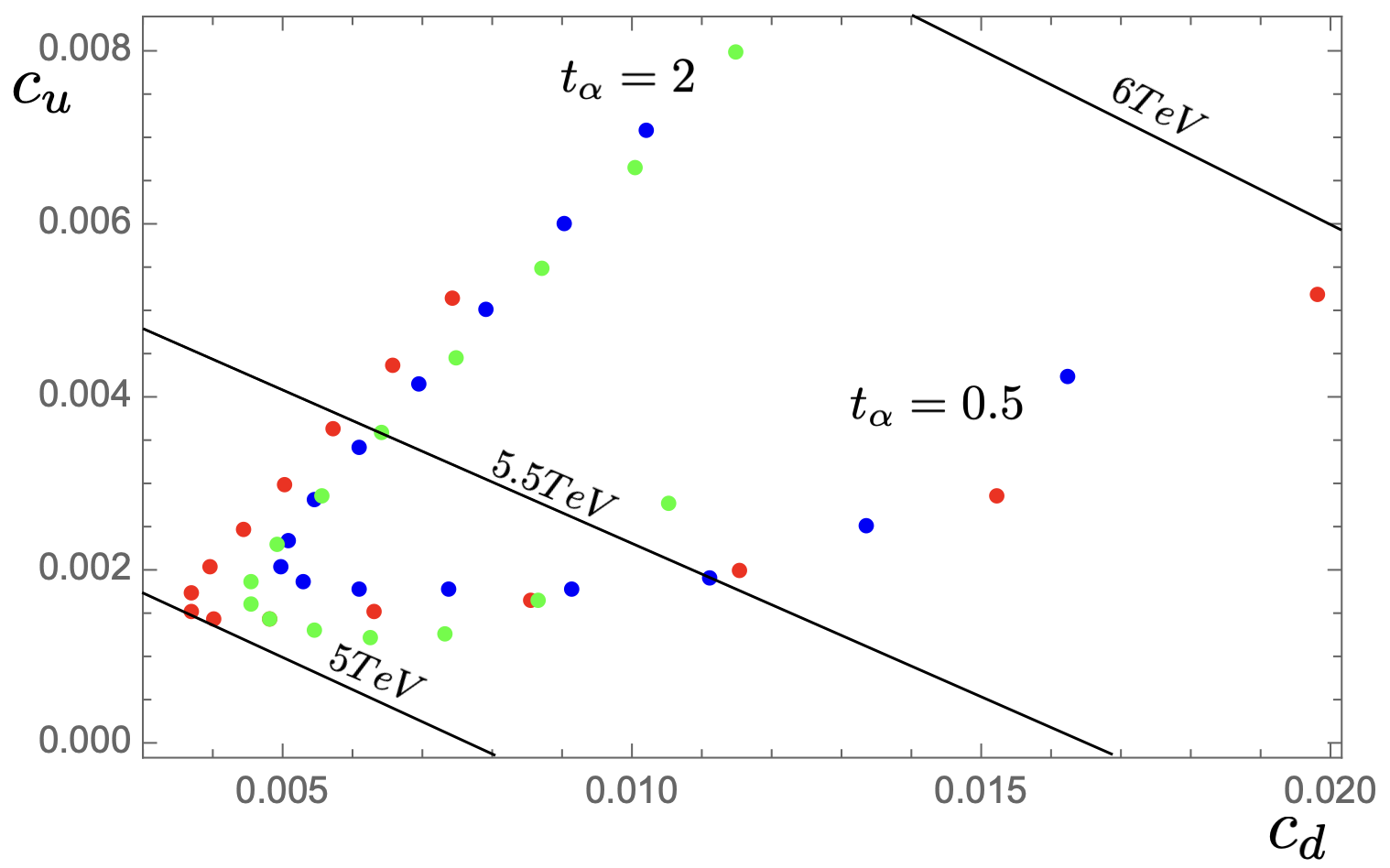}
\caption{\small  Lower limits on $m_{Z_{23}}$ from the Drell-Yan process in the $(c_u, c_d)$ plane, compared with the model predictions for $c_{u,d}$ at equidistant  values of $t_\alpha$ from $0.5$ to $2$ for $b=0.25 (green), b=1 (blue), b=4 (red)$ }
\label{fig:c_ud}
\end{figure}

The data collected by ATLAS \cite{ATLAS:2019erb} and CMS \cite{CMS:2021ctt}  during Run 2 at LHC do not show any evidence of a signal, so we use a $95\%$ CL bound on  $\sigma_{l^+l^-} $. By deducing $g^f_{L,R}$ for every chiral fermion $f$ from the identification of Eq.~(\ref{gVA}) with Eq.~(\ref{Dmu-phys})  for the  $Z_{23}$ part, one gets the lower limits on  $m_{Z_{23}}$ shown in Fig. \ref{fig:c_ud} in the $(c_u, c_d)$ plane and compared with the model predictions  for $c_{u,d}$. The data inputs are summarised in Table~\ref{Tab:data} for three relevant values of $m_{Z_{23}}$~\cite{ATLAS:2019erb, CMS:2021ctt, Baker:2024xwh
}.
The narrow width approximation is justified since, for any $t_\alpha$ and $b$ in the range of interest, $\frac{\Gamma}{m}|_{Z_{23}}$ is below $3\%$. In the model under consideration proton-proton collisions into jet pairs are not competitive with Drell-Yan.

\subsection{Flavour changing effects}

The different couplings in Eq.s~(\ref{Dmu-phys},{\ref{Dmu-Z}) of $Z_{23}^{(')}$ and of the $Z$ itself to the third versus the first two fermion generations lead  to  Flavour Changing (FC) interactions. 
After going from the interaction basis, $f^{(0)}, f=u,d,e$, to the mass eigenstates, $f_L=U^f_L f_L^{(0)}$,   $f_R\approx f_R^{(0)}$, one gets
  \begin{equation}
 \mathcal{L}^{(FC)} = [ g' (a_{23}-\frac{1}{t_\alpha})Z_{23\mu} + g_3 Z'_{23\mu} +
 s_W g' \frac{m_Z^2}{m_{Z_{23}}^2} (a_{23}^2-\frac{a_{23}}{t_\alpha} +a_3)
 Z_{\mu} ] J^\mu,
  \end{equation}
  where
    \begin{equation}
    J^\mu=
 \frac{1}{6}J_u^\mu + \frac{1}{6}J_d^\mu -\frac{1}{2}J_e^\mu,\quad\quad
 J_f^\mu = \Sigma_{i\neq j} [U^f_L ]_{i3}[U^f_L ]_{j3}^* \bar{f}_{Li}\gamma^\mu f_{Lj}.
 \end{equation}
In turn the tree level exchanges of $Z, Z_{23}^{(')}$ give rise to a number of effective FC four-fermion interactions, all weighted by $g'^2/m_{Z_{23}}^2$, whose effects are discussed below in some  relevant, although not exclusive, examples.

\subsubsection{$\Delta F=2 $  transitions}

For the $\Delta F=2$ transitions, from the $Z_{23}$ and  $Z_{23}'$ exchanges, one gets
\begin{equation}
\mathcal{L}_{eff}(\Delta F=2)= -C_1^{B_s} (\bar{s}\gamma^\mu b)^2
-C_1^{B_d} (\bar{d}\gamma^\mu b)^2 -C_1^{D} (\bar{u}\gamma^\mu c)^2
\end{equation}
where, using Eq.~(\ref{a3}),
\begin{equation}
C_1^{B_s} = \frac{g'^2}{36 m_{Z_{23}}^2}
([U^d_L]_{s3}[U^{d}_L]_{b3}^* )^2 [(a_{23}-\frac{1}{t_\alpha})^2 + a_3]
\end{equation}
and similarly for $C_1^{B_d}$ and $C_1^{D}$ with the only replacement of the appropriate $U_L$ matrix elements.

Knowing that  $U_L^{u,d}\approx V_{CKM}$  (see Section~\ref{Mod}), one can estimate the size of these effects by considering $U_L^u = \bold{1}$ in the $C_1^{B_{d,s}}$ case or $U_L^d = \bold{1}$ in the $C_1^{D}$ case. The constraint on $m_{Z_{23}}$ is shown in Fig.~\ref{fig:DeltaB=2}, using the current bound on $|C_1^{B_s}|$\cite{Bona:2024bue}. Somewhat weaker bounds are obtained from $C_1^{B_d}$ and $C_1^{D}$.

\subsubsection{$b\rightarrow s   l^+l^-, l=e,\mu$ }
\begin{figure}[t]
\centering
\includegraphics[clip,width=.55\textwidth]{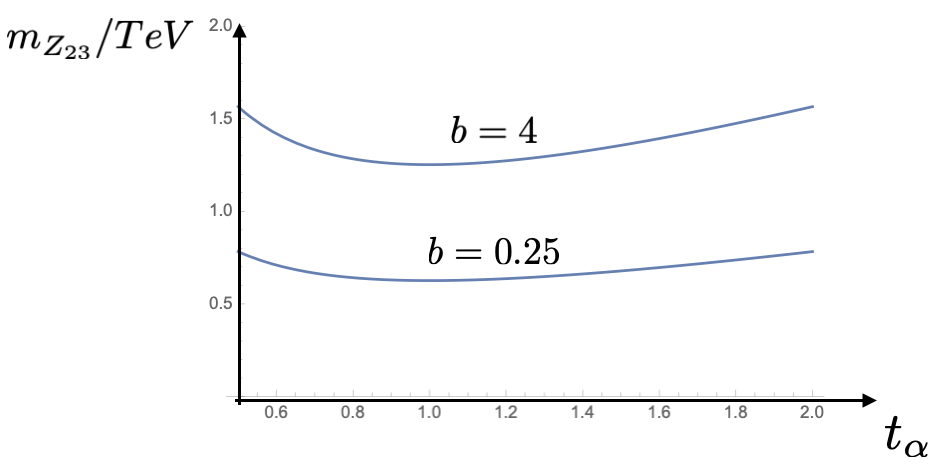}
\caption{\small  Lower bound on  $m_{Z_{23}}$ from $B_s$ meson mixing for $b=0.25$ and $b=4$ and  $|[U^d_L]_{s3}[U^{d}_L]_{b3}^* /V_{tb}V_{ts}^*|=1$. Intermediate values of $b$ give intermediate values of the bound. }
\label{fig:DeltaB=2}
\end{figure}

The possible existence of new physics in $b\rightarrow s   l^+l^-, l=e,\mu$  is traditionally analysed in terms of the effective Lagrangian
\begin{equation}
\mathcal{L}_{eff}(b\rightarrow s   l^+l^-)=\frac{4 G_F}{\sqrt{2}}\frac{e^2}{16\pi^2}V_{ts}^*V_{tb}(C_9 \mathcal{O}_9 + C_{10} \mathcal{O}_{10})
\end{equation}
where
\begin{equation}
 \mathcal{O}_9 = (\bar{s}_L\gamma^\mu b_L)(\bar{l}\gamma_\mu l)           ,\quad\quad  
 \mathcal{O}_{10} = (\bar{s}_L\gamma^\mu b_L)(\bar{l}\gamma_\mu\gamma_5 l).
\end{equation}
 A fit of several observables with one single operator at a time, gives $C_{10}$ consistent with $0$, whereas, although with significant theoretical uncertainties, $C_9=
-0.78\pm 0.21$~\cite{Greljo:2022jac}.

These operators are generated, universally for $e$ and $\mu$, by the exchange of the $Z_{23}$ and of the $Z$ itself.
One finds
\begin{equation}
 \mathcal{L}_{eff}(Z_{23}, b\rightarrow s   l^+l^-)  = - \frac{g'^2}{24 m_{Z_{23}}^2}[U^d_L]_{s3}[U^{d}_L]_{b3}^* 
 (a_{23}-\frac{1}{t_\alpha})[(t_\alpha-\frac{2}{t_\alpha})  \mathcal{O}_{9} + t_\alpha \mathcal{O}_{10} ] ,
 \end{equation}
 \begin{equation}
  \mathcal{L}_{eff}(Z, b\rightarrow s   l^+l^-) = - \frac{g'^2}{24 m_{Z_{23}}^2}[U^d_L]_{s3}[U^{d}_L]_{b3}^* 
  (a_{23}^2- \frac{1}{t_\alpha} a_{23}+a_3)    [(1-4s_W^2)\mathcal{O}_9 +  \mathcal{O}_{10} ]
\end{equation}
Their sum gives $C_{10}$  proportional to the combination
\begin{equation}
a_{23}^2 + a_3 -\frac{1}{t_\alpha}a_{23} + t_\alpha a_{23} -1,
\end{equation}
which vanishes identically for any value of $b$ and $t_\alpha$. For $C_9$, neglecting the contribution from $Z$-exchange proportional to $(1-4s_W^2)$, one finds
\begin{equation}
C_9 = -\frac{\sqrt{2}}{G_F} \frac{\pi^2}{6 c_W^2} \frac{1}{m_{Z_{23}}^2} \frac{[U^d_L]_{s3}[U^{d}_L]_{b3}^* }{V_{ts}^*V_{tb}}
(a_{23}-\frac{1}{t_\alpha})(t_\alpha -\frac{2}{t_\alpha})
\end{equation}
The experimental interval quoted above can be reproduced only for values of $m_{Z_{23}}$ at or below the bounds shown in Fig.~\ref{fig:DeltaB=2} from $B$-meson mixing.

\subsubsection{$\mu\rightarrow 3 e$ and $\tau\rightarrow 3\mu$}

\begin{figure}[t]
\centering
\includegraphics[clip,width=.55\textwidth]{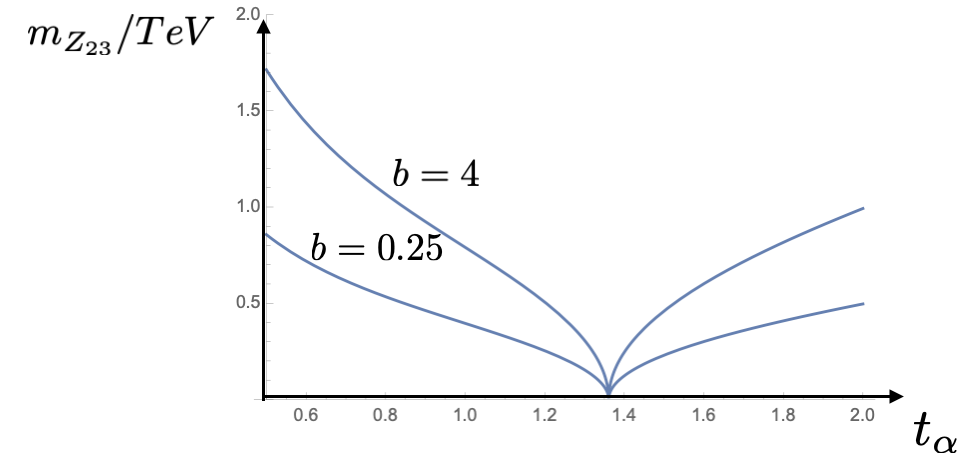}
\caption{\small  Lower bound on  $m_{Z_{23}}$ from $BR(\tau\rightarrow 3 \mu) < 2\cdot 10^{-8}$ for $b=0.25$ and $b=4$ and  $|[U^e_L]_{\mu 3}[U^{e}_L]_{\tau 3}^* /V_{tb}V_{ts}^*|=1$. Intermediate values of $b$ give intermediate values of the bound. }
\label{fig:tau->3mu}
\end{figure}

In total analogy with the previous cases, the exchanges of of the $Z_{23}$ and of the $Z$ give rise to $\mu\rightarrow 3 e$ and to  $\tau\rightarrow 3 \mu$  amplitudes described by the effective Lagrangian
\begin{eqnarray}
  \mathcal{L}_{eff} &=&  - \frac{g'^2}{4 m_{Z_{23}}^2}( [U^e_L]_{\mu 3}[U^{e}_L]_{\tau 3}^* \bar{\mu}_L\gamma_\mu \tau_L +
  [U^e_L]_{e 3}[U^{e}_L]_{\mu 3}^* \bar{e}_L\gamma_\mu \mu_L) \cdot \nonumber \\
&& \cdot (a_L(\bar{e}_L\gamma_\mu e_L + \bar{\mu}_L\gamma_\mu \mu_L) +
  a_R(\bar{e}_R\gamma_\mu e_R + \bar{\mu}_R\gamma_\mu \mu_R) )
\end{eqnarray}
where
\begin{eqnarray}
 &&  a_L= \frac{1}{t_\alpha} (a_{23}- \frac{1}{t_\alpha} ) + (1-2s_W^2) (a_{23}(a_{23}- \frac{1}{t_\alpha} )+a_3)
  \nonumber \\
&&  a_R= \frac{1}{t_\alpha} (1- t_\alpha^2)(a_{23}- \frac{1}{t_\alpha} ) -2s_W^2 (a_{23}(a_{23}- \frac{1}{t_\alpha} )+a_3)
\end{eqnarray}
One can have a feeling of the size of these effects by taking $U_L^e =V_{CKM}^+$, i.e. $U_L^e =U_L^d $ and $U_L^u = \bold{1}$. In this case from $BR(\tau\rightarrow 3 \mu) < 2\cdot 10^{-8}$ one obtains the lower bound on $m_{Z_{23}}$ shown in Fig.~\ref{fig:tau->3mu}, with a similar, slightly weaker bound from $BR(\mu\rightarrow 3 e) <  10^{-12}$.

\section{Loop effects due to light-heavy fermion mixings}

As already discussed in Ref.~\cite{Barbieri:2023qpf}, one loop effects induced by light-heavy fermion mixings can give rise to significant interactions. The general structure of these mixings is recalled in Appendix A. Here I refine the discussion of the leptonic dipole operators, potentially relevant for the electric and magnetic dipole moments as for the radiative Lepton Flavour Changing interactions\footnote{
As pointed out in Ref.~\cite{Barbieri:2023qpf} an important effect of the  light-heavy fermion mixings is the $\Delta S=2$ amplitude  resulting from the 
box diagram with heavy fermions and the (SM-like) $H_d$ field, which gives the bound $M_{23}>80~TeV\cdot \sqrt{Im ( y^d_{21} y^{d*}_{11} /10^{-1})^2}$. This bound, however,  is not incompatible with $M_{23}= m_{Z_{23}}/g' \epsilon_{\phi, \chi}\approx 
140~TeV (m_{Z_{23}}/5 TeV) (10^{-1}/\epsilon_{\phi, \chi})$.
}.

The leading corrections to the leptonic dipole operators come from the exchange of $Z_{23}$ with its couplings to the light-heavy charged leptons. Before mixing the $Z_{23}$-couplings to leptons can be written as
\begin{equation}
\mathcal{L}_{Z_{23}} = \Sigma_{Q} g' a_Q Z_{23\mu}(\bar{e}^{(0)}_L\gamma^\mu Q_{e_L} e^{(0)}_L + \bar{E}^{(0)}_L\gamma^\mu Q_{E} E^{(0)}_L)  + L\leftrightarrow R
\end{equation}
where $Q= Y^{[3]}, 1/2(B-L)^{[12]}, T_{3R}^{[12]}$ are $(3\times 3)$ diagonal matrices, specified in Table \ref{Table:VLFA}, and the coefficients  $a_Q$ are given in Eq. \ref{Dmu-phys}. After mixing, based on the transformations~(\ref{mixings}), the light-heavy lepton couplings emerge
\begin{equation}
\mathcal{L}_{Z_{23}}^{light-heavy} = \Sigma_{Q} g' a_Q Z_{23\mu}(\bar{e}_L\gamma^\mu \hat{\epsilon}_L(Q) E_L  + L\leftrightarrow R) + h.c.,
\end{equation}
where 
\begin{equation}
 \hat{\epsilon}_{L,R}(Q)= Q_{e_{L,R}}\epsilon^e_{L,R}- \epsilon^e_{L,R} Q_E,
\end{equation}
and the explicit expressions for $\epsilon^e_{L,R}$ are given in Appendix A. As a consequence the one loop dipole operators to leading order in $m^2_{Z_{23\mu}}/M^2$ are\cite{Kannike_2012,Freitas_2014}

\begin{equation}
\mathcal{L}^{dip} = \frac{e}{32\pi^2}\frac{g'^2}{m^2_{Z_{23\mu}}}
\Sigma_{Q, Q'} a_Q a_{Q'}
(\bar{e}_L\sigma^{\mu\nu}  \hat{\epsilon}_L(Q) M  \hat{\epsilon}_R^{\dagger}(Q') e_R) F_{\mu\nu}.
\label{dip}
\end{equation}

By explicit calculation, out of the nine possible matrices $  \hat{\epsilon}_L(Q) M  \hat{\epsilon}_R^{\dagger}(Q') $, to a sufficient approximation only four are non vanishing:
\begin{equation}
\hat{\epsilon}_L(Y^{[3]})M  \hat{\epsilon}_R^{\dagger}(Y^{[3]}) = \hat{\epsilon}_L(T_{3R}^{[12]})M  \hat{\epsilon}_R^{\dagger}(Y^{[3]}) =
\frac{v_1}{4}
\begin{pmatrix}
-y_1^\sigma {y}_{1\alpha}^e \hat{y}_{\alpha 3}^{\phi_e} \epsilon_\phi \epsilon_\sigma & {y}_{1\alpha}^e y^{\phi_e}_{\alpha 2}\epsilon_\phi & -y_{12}^e y_2^{\chi_e}\epsilon_{\chi^l} \\
-y_1^\sigma {y}_{2\alpha}^e \hat{y}_{\alpha 3}^{\phi_e} \epsilon_\phi \epsilon_\sigma & {y}_{2\alpha}^e y^{\phi_e}_{\alpha 2}\epsilon_\phi &  -y_{22}^e y_2^{\chi_e}\epsilon_{\chi^l}\\
0 &0 &  0
\end{pmatrix}
\end{equation}
and
\begin{equation}
\hat{\epsilon}_L(Y^{[3]})M  \hat{\epsilon}_R^{\dagger}(T_{3R}^{[12]}) = \hat{\epsilon}_L(T_{3R}^{[12]})M  \hat{\epsilon}_R^{\dagger}(T_{3R}^{[12]}) =
-\frac{v_1}{4}
\begin{pmatrix}
-y_1^\sigma {y}_{1\alpha}^e \hat{y}_{\alpha 3}^{\phi_e} \epsilon_\phi \epsilon_\sigma & {y}_{1\alpha}^e y^{\phi_e}_{\alpha 2}\epsilon_\phi & 0 \\
-y_1^\sigma {y}_{2\alpha}^e \hat{y}_{\alpha 3}^{\phi_e} \epsilon_\phi \epsilon_\sigma & {y}_{2\alpha}^e y^{\phi_e}_{\alpha 2}\epsilon_\phi &  0\\
0 &0 &  0
\end{pmatrix}
\end{equation}
By comparison with the Yukawa coupling matrix for the charged leptons $Y_e$ - Eq.~(\ref{eq:YuA}) with $u \rightarrow e$ - the effective dipole interaction, Eq.~(\ref{dip}) can be written in the form
\begin{equation}
\mathcal{L}^{dip} = \frac{e v_1}{32\pi^2}\frac{g'^2}{m^2_{Z_{23\mu}}} 
\bar{e}_L 
(a_Y Y_e + \tilde{a}_Y \tilde{Y}_e)
e_R F_{\mu\nu}.
\label{dip2}
\end{equation}
where $a_Y, \tilde{a}_Y$ are real coefficients depending on $t_\alpha, b$, as the $a_Q$     are, and $\tilde{Y}_e$ has non vanishing elements only on the third column, of similar size to the corresponding elements in ${Y}_e$. 

Since $a_Y, \tilde{a}_Y$ are of order unity, after going to the physical basis, $e_L \rightarrow U^e_L e_L, e_R  \rightarrow U^{e\dagger}_R e_R$, the term proportional to $Y_e $ gives no significant effect irrespective of the values of $U^e_{L,R}$. The same is  largely true for  the term proportional to $\tilde{Y}_e $  as well, if one considers the typical  size of the off diagonal elements of $U^e_{L,R}$ as it arises from the actual form of $Y_e $.

\section{Summary}

The role of suitable flavour symmetries in keeping under control the effects of flavour-changing higher dimensional operators has often been invoked, thus allowing for them to be weighted by a scale not far from the TeV. These  symmetries may be related to the structure of the Yukawa couplings and may perhaps arise accidentally, thus strongly motivating the search for new phenomena by precision experiments in several different areas before the commissioning of a much needed higher energy collider.

The  implementation of this picture in an explicit 4d gauge theory has allowed to analyse in detail the full phenomenology at the lightest possible new scale, the mass $m_{Z_{23}}$ of the lightest neutral gauge boson. The results are shown in Fig.s \ref{fig:deltam_Z} to \ref{fig:tau->3mu} in some  significant, although not exclusive, examples.
Two new parameters enter in ElectroWeak and high-$p_T$ precision physics. Flavour physics observables depend as well on the matrix elements $[U^f_L ]_{i3}[U^f_L ]_{j3}^*, f=u,d,e;
i,j=1,2,3$, predicted of similar size but not identical to  $[V_{CKM}]_{t i}[V_{CKM}]_{t j}^*, i,j=d,s,b$ (and subject to the constraint $U_L^u U_L^{d+} = V_{CKM}$). 
In judging of the apparent dominance of the current constraints from the corrections to the $Z$ mass, Fig.~\ref{fig:deltam_Z},
and from the Drell-Yan process, Fig.~\ref{fig:c_ud}, relative to the current flavour constraints, Fig.~\ref{fig:DeltaB=2},\ref{fig:tau->3mu},
one has to keep in mind the specific identification of these matrix elements in the various explored flavour observables.
A peculiar result, irrespective of the values of these matrix elements, is the vanishing of the coefficient $C_{10}$ in $\mathcal{L}_{eff}(b\rightarrow s   l^+l^-)$.

 Increased precision in the mid term future of ElectroWeak Physics, the Drell-Yan process  and flavour observables at HL-LHC and at various flavour experiments can in each of these cases find evidence for motivated new  physics in the MultiTeV. The emergence of a convincing deviation from the SM in any of these observables would motivate an overall fit along the lines outlined above.

\section*{Acknowledgments}

I am indebted to Gino Isidori for the collaboration  on Ref.~\cite{Barbieri:2023qpf} and for many useful discussions on the general subject and to Riccardo Torre for his help on Section 5.2.

\section*{Appendix A: Light-heavy fermion mixing}
\label{app:A}

The Lagrangian~(\ref{LYA}) and its analogue for the down quarks and the charged leptons gives rise, once the scalar fields get their vev, to a $(6\times 6)$ mass matrix
\begin{equation}
{\mathcal L}^{(m_f)}  = \begin{pmatrix}
\bar{f}_L^{(0)}&\bar{F}_L^{(0)}
\end{pmatrix}
\begin{pmatrix}
m^f& \Delta_L^f \\
\Delta_R^f &{M^f}
\end{pmatrix}
\begin{pmatrix}
f_R^{(0)} \\
F_R^{(0)}
\end{pmatrix}\,,\quad f=u,d,e
\label{massmatrix}
\end{equation}
\begin{equation}
m^f=v_f
\begin{pmatrix}
0&0&0 \\
0&0&0 \\
0&0& y_3^f
\end{pmatrix},\quad
\Delta_L^f =v_f
\begin{pmatrix}
y^f_{11}&y^f_{12}  & 0\\
y^f_{21}& y^f_{22}& 0  \\
0&0&0
\end{pmatrix},\quad
\Delta_R^f =
\begin{pmatrix}
0& y^{\phi_f}_{1 2}  \langle \phi \rangle & 0 \\
0& y^{\phi_f}_{2 2}  \langle \phi \rangle & y^{\chi_f}_{2} \langle \chi^f \rangle  \\
y_1^\sigma  \langle \sigma \rangle  & 0 &   0
\end{pmatrix},
\label{eq:DeltaM}
\end{equation}
all perturbative matrices relative to 
\begin{equation}
M^f=
\begin{pmatrix}
M_{[23]}&0&  \hat{y}^{\phi_f}_{1 3}  \langle \phi \rangle  \\
0 &M_{[23]}&   \hat{y}^{\phi_f}_{2 3}  \langle \phi \rangle \\
 y^{\phi_f * }_{1 3}  \langle \phi \rangle&  y^{\phi_f * }_{2 3}  \langle \phi \rangle &M_{[12]}
\end{pmatrix},
\end{equation}
To leading order in $1/M^f$ the Left Handed and Right Handed rotations that eliminate the light-heavy terms are
 \begin{equation}
 \begin{pmatrix}
f_R^{(0)} \\
F_R^{(0)}
\end{pmatrix} =
\begin{pmatrix}
\bold{1}&\epsilon_R^f\\
-\epsilon^{f\dagger}_R&\bold{1}
\end{pmatrix}
\begin{pmatrix}
f_R\\
F_R
\end{pmatrix}\,,
\qquad 
 \begin{pmatrix}
f_L^{(0)} \\
F_L^{(0)}
\end{pmatrix} =
\begin{pmatrix}
\bold{1}&\epsilon_L^f\\
-\epsilon_L^{f\dagger} &\bold{1}
\end{pmatrix}
\begin{pmatrix}
f_L\\
F_L
\end{pmatrix}\,,
\label{mixings}
\end{equation}
where
\begin{equation}
\epsilon_L^f =v_f
\begin{pmatrix}
\frac{y^f_{11}}{M_{[23]}}  & \frac{y^f_{12}}{M_{[23]}} & -y^f_{1\alpha}\hat{y}^{\phi_f}_{\alpha 3}\frac{\epsilon_\phi}{M_{[12]}} \\
\frac{y^f_{21}}{M_{[23]}} & \frac{y^f_{22}}{M_{[23]}} & -y^f_{2\alpha}\hat{y}^{\phi_f}_{\alpha 3}\frac{\epsilon_\phi}{M_{[12]}}  \\
0&0 & 0
\end{pmatrix}, \quad
\epsilon_R^{f\dagger} = 
\begin{pmatrix}
-y_1^\sigma \hat{y}_{13}^{\phi_f}\epsilon_\phi \epsilon_\sigma & y^{\phi_f}_{12}\epsilon_\phi & 0\\
-y_1^\sigma \hat{y}_{13}^{\phi_f}\epsilon_\phi \epsilon_\sigma & y^{\phi_f}_{12}\epsilon_\phi & y_2^{\chi^f}\epsilon_{\chi^f} \\
y_1^\sigma\epsilon_\sigma &\approx 0 & \approx 0
\end{pmatrix}
   \end{equation}

\bibliographystyle{JHEP}
\bibliography{references-2024}

\providecommand{\href}[2]{#2}\begingroup\raggedright\begin{thebibliography}{10}

\bibitem{Barbieri:2023qpf}
R.~Barbieri and G.~Isidori, \emph{{Minimal flavour deconstruction}},
  \href{https://doi.org/10.1007/JHEP05(2024)033}{\emph{JHEP} {\bfseries 05}
  (2024) 033}, [\href{https://arxiv.org/abs/2312.14004}{{\ttfamily
  2312.14004}}].

\bibitem{Bona:2024bue}
M.~Bona et~al., \emph{{Overview and theoretical prospects for CKM matrix and CP
  violation from the UTfit Collaboration}},
  \href{https://doi.org/10.22323/1.457.0007}{\emph{PoS} {\bfseries WIFAI2023}
  (2024) 007}.

\bibitem{Barbieri:2011ci}
R.~Barbieri, G.~Isidori, J.~Jones-Perez, P.~Lodone and D.~M. Straub,
  \emph{{$U(2)$ and Minimal Flavour Violation in Supersymmetry}},
  \href{https://doi.org/10.1140/epjc/s10052-011-1725-z}{\emph{Eur. Phys. J. C}
  {\bfseries 71} (2011) 1725},
  [\href{https://arxiv.org/abs/1105.2296}{{\ttfamily 1105.2296}}].

\bibitem{Davighi:2022bqf}
J.~Davighi, G.~Isidori and M.~Pesut, \emph{{Electroweak-flavour and
  quark-lepton unification: a family non-universal path}},
  \href{https://doi.org/10.1007/JHEP04(2023)030}{\emph{JHEP} {\bfseries 04}
  (2023) 030}, [\href{https://arxiv.org/abs/2212.06163}{{\ttfamily
  2212.06163}}].

\bibitem{Davighi:2023iks}
J.~Davighi and G.~Isidori, \emph{{Non-universal gauge interactions addressing
  the inescapable link between Higgs and flavour}},
  \href{https://doi.org/10.1007/JHEP07(2023)147}{\emph{JHEP} {\bfseries 07}
  (2023) 147}, [\href{https://arxiv.org/abs/2303.01520}{{\ttfamily
  2303.01520}}].

\bibitem{FernandezNavarro:2023rhv}
M.~Fern\'andez~Navarro and S.~F. King, \emph{{Tri-hypercharge: a separate
  gauged weak hypercharge for each fermion family as the origin of flavour}},
  \href{https://doi.org/10.1007/JHEP08(2023)020}{\emph{JHEP} {\bfseries 08}
  (2023) 020}, [\href{https://arxiv.org/abs/2305.07690}{{\ttfamily
  2305.07690}}].

\bibitem{Davighi:2023evx}
J.~Davighi and B.~A. Stefanek, \emph{{Deconstructed Hypercharge: A Natural
  Model of Flavour}},  \href{https://arxiv.org/abs/2305.16280}{{\ttfamily
  2305.16280}}.

\bibitem{Covone:2024elw}
S.~Covone, J.~Davighi, G.~Isidori and M.~Pesut, \emph{{Flavour Deconstructing
  the Composite Higgs}},  \href{https://arxiv.org/abs/2407.10950}{{\ttfamily
  2407.10950}}.

\bibitem{deBlas:2021wap}
J.~de~Blas, M.~Ciuchini, E.~Franco, A.~Goncalves, S.~Mishima, M.~Pierini
  et~al., \emph{{Global analysis of electroweak data in the Standard Model}},
  \href{https://doi.org/10.1103/PhysRevD.106.033003}{\emph{Phys. Rev. D}
  {\bfseries 106} (2022) 033003},
  [\href{https://arxiv.org/abs/2112.07274}{{\ttfamily 2112.07274}}].

\bibitem{Altarelli:1991fk}
G.~Altarelli, R.~Barbieri and S.~Jadach, \emph{{Toward a model independent
  analysis of electroweak data}},
  \href{https://doi.org/10.1016/0550-3213(92)90376-M}{\emph{Nucl. Phys. B}
  {\bfseries 369} (1992) 3--32}.

\bibitem{ParticleDataGroup:2022pth}
{\scshape Particle Data Group} collaboration, R.~L. Workman et~al.,
  \emph{{Review of Particle Physics}},
  \href{https://doi.org/10.1093/ptep/ptac097}{\emph{PTEP} {\bfseries 2022}
  (2022) 083C01}.

\bibitem{CDF:2022hxs}
{\scshape CDF} collaboration, T.~Aaltonen et~al., \emph{{High-precision
  measurement of the $W$ boson mass with the CDF II detector}},
  \href{https://doi.org/10.1126/science.abk1781}{\emph{Science} {\bfseries 376}
  (2022) 170--176}.

\bibitem{Carena:2004xs}
M.~Carena, A.~Daleo, B.~A. Dobrescu and T.~M.~P. Tait, \emph{{$Z^\prime$ gauge
  bosons at the Tevatron}},
  \href{https://doi.org/10.1103/PhysRevD.70.093009}{\emph{Phys. Rev. D}
  {\bfseries 70} (2004) 093009},
  [\href{https://arxiv.org/abs/hep-ph/0408098}{{\ttfamily hep-ph/0408098}}].

\bibitem{Accomando:2010fz}
E.~Accomando, A.~Belyaev, L.~Fedeli, S.~F. King and C.~Shepherd-Themistocleous,
  \emph{{Z' physics with early LHC data}},
  \href{https://doi.org/10.1103/PhysRevD.83.075012}{\emph{Phys. Rev. D}
  {\bfseries 83} (2011) 075012},
  [\href{https://arxiv.org/abs/1010.6058}{{\ttfamily 1010.6058}}].

\bibitem{ATLAS:2019erb}
{\scshape ATLAS} collaboration, G.~Aad et~al., \emph{{Search for high-mass
  dilepton resonances using 139 fb$^{-1}$ of $pp$ collision data collected at
  $\sqrt{s}=$13 TeV with the ATLAS detector}},
  \href{https://doi.org/10.1016/j.physletb.2019.07.016}{\emph{Phys. Lett. B}
  {\bfseries 796} (2019) 68--87},
  [\href{https://arxiv.org/abs/1903.06248}{{\ttfamily 1903.06248}}].

\bibitem{CMS:2021ctt}
{\scshape CMS} collaboration, A.~M. Sirunyan et~al., \emph{{Search for resonant
  and nonresonant new phenomena in high-mass dilepton final states at $
  \sqrt{s} $ = 13 TeV}},
  \href{https://doi.org/10.1007/JHEP07(2021)208}{\emph{JHEP} {\bfseries 07}
  (2021) 208}, [\href{https://arxiv.org/abs/2103.02708}{{\ttfamily
  2103.02708}}].

\bibitem{Baker:2024xwh}
M.~J. Baker, T.~Martonhelyi, A.~Thamm and R.~Torre, \emph{{A Simplified Model
  of Heavy Vector Singlets for the LHC and Future Colliders}},
  \href{https://arxiv.org/abs/2407.11117}{{\ttfamily 2407.11117}}.

\bibitem{Greljo:2022jac}
A.~Greljo, J.~Salko, A.~Smolkovi\v{c} and P.~Stangl, \emph{{Rare b decays meet
  high-mass Drell-Yan}},
  \href{https://doi.org/10.1007/JHEP05(2023)087}{\emph{JHEP} {\bfseries 05}
  (2023) 087}, [\href{https://arxiv.org/abs/2212.10497}{{\ttfamily
  2212.10497}}].

\bibitem{Kannike_2012}
K.~Kannike, M.~Raidal, D.~M. Straub and A.~Strumia, \emph{Anthropic solution to
  the magnetic muon anomaly: the charged see-saw},
  \href{https://doi.org/10.1007/jhep02(2012)106}{\emph{Journal of High Energy
  Physics} {\bfseries 2012} (Feb., 2012) }.

\bibitem{Freitas_2014}
A.~Freitas, J.~Lykken, S.~Kell and S.~Westhoff, \emph{Testing the muon g-2
  anomaly at the lhc},
  \href{https://doi.org/10.1007/jhep05(2014)145}{\emph{Journal of High Energy
  Physics} {\bfseries 2014} (May, 2014) }.

\end{thebibliography}\endgroup
\end{document}